# Lake Titicaca—Physics of an Inherited Hydropower Macroproject Proposal

Richard B. Cathcart
Geographos
Burbank, California
USA
e-mail: rbcathcart@charter.net

Alexander A. Bolonkin
C & R
Brooklyn, New York
USA
e-mail: abolonkin@juno.com

**Abstract**

Shared almost evenly by Peru and Bolivia, Lake Titicaca is situated on an Altiplano endorheic region of the northern Andes Mountains. Rio Desaguadero is the lake's only outlet. From 1908, several macro-engineers speculated on the creation of a second, completely artificial, outlet for Lake Titicaca's freshwater. Here we reconsider several 20$^{th}$ Century macroproject proposals, with the goal of examining and enhancing this technically interesting South American 21$^{st}$ Century Macro-engineering inheritance.

**1. Introduction**

Lake Titicaca is, probably, a Quaternary Period remnant of the Mantaro mega-lake that filled the Basin to a maximum elevation of ~4,000 m (Oncken, 2006). Nowadays, Lake Titicaca has a semi-regulated surface altitude of ~3810 m. Its freshwater volume is ~930 km$^3$ (UNESCO, 2003). During an extremely dry period between circa 4000 BC to circa 3000 BC, Lake Titicaca was ~75 m below its present-day altitude; by circa 2500 BC it began to fill again and rise to its current elevation (Baker, 2005). The presently existing higher elevation of the lake's free surface has submerged ancient artifacts and settlements (Calaway, 2005). Indeed, the landscape surrounding the present-day Lake Titicaca indicates an ancient anthropogenic component—terraces, raised fields, sunken gardens and irrigated pastures (Morris, 2004).

Lake Titicaca constitutes a freshwater resource for approximately 2.5-3.0 million persons residing in Peru and Bolivia. However, about 95% of the freshwater removed from Lake Titicaca leaves by way of evaporation, while the remaining 5% is comprised of a discharge through the Rio Desaguadero amounting to ~20-35 m$^3$/second. During 1996, the two riparian states organized the Binational Autonomous Authority for Lake Titicaca (ALT). Since 2001, a dam near the International Bridge at the headwater of the Rio Desaguadero has been equipped with floodgates to provide a stable anthropogenic hydraulic capacity for Lake Titicaca. According to the ALT Master Plan, the water level of Lake Titicaca ideally ought to be maintained at range of between 3808 and 3811 m above sea level. An increase in the human population around Lake Titicaca is forecast and freshwater usage will inevitably increase (Revollo, 2001). A future resource for the needed freshwater may become possible with the reduction of natural evaporation from the lake's ~8,400 km$^2$ surface. In similar arid climates elsewhere as, for example, at Lake Eyre in Australia, a 50% reduction of free water surface evaporation could be achieved through deliberate spreading of an ultra-thin layer of organic molecules (in powder form) such as proposed by Robert Neville O'Brien in USA Patent 6303133, issued 16

October 2001. Under the sunny conditions prevailing at Lake Titicaca, after application biodegradation will occur in about 2.5 days; it will be technically necessary, then, for ~100 kg of O'Brien's powder to be broadcast three times per week, or 12 times per month—144 times per year.

Today, Peru's ~28,000,000 citizens obtain ~80% of their power from hydroelectric generating facilities. The nation's total operational base-load is nearly 6000 MW. (The country's largest hydropower installation is the Mantaro Complex in southern Peru. About 20% of Peru's hydroelectricity is generated there, nearly 1008 MW. The Mantaro Complex—built from 1973 to 1985—utilizes a ~1,000 m drop which occurs at the Mantaro River's first great bend as it flows into the Basin of the Amazon River.) Peru has untapped hydropower resources on the eastern slopes of the Andes Mountains while thermal power stations are mostly concentrated along the densely populated coast affected directly by the Pacific Ocean.

**2. Hydropower from Lake Titicaca for the Coast**

While the National Map of Peru was commenced on 10 May 1921 by plane table and alidade methods, it remains incomplete at the start of the 21$^{st}$ Century (Mugnier, 2006). Accurate topographic mapping is, of course, essential to the planning of any macroproject, but especially a macroproject that relies on freshwater's long distance fall from the Andes Mountains to the Pacific Ocean!

Charles Reginald Enock (1868-1970), a British explorer of Peru, was first to explore and cursorily map the resources of Lake Titicaca and its surrounding region (Markham, 1905). He suggested a 120 km-long tunnel could convey the lake's liquid contents to the populated coastal region adjacent to the Pacific Ocean and generate hydropower simultaneously (Enock, 1908). A vertical drop to sea level of, say, 20 m$^3$/second from an elevation of 3810 m could theoretically produce 640 MW, close to 10% of Peru's existing installed electricity generation infrastructure! Although we cannot be certain, we suspect C.R. Enock may have been inspired by the North American engineer Alexis Von Schmidt (1821-1906) who, from 1865 onwards, proposed and promoted a freshwater aqueduct to the City and County of San Francisco drawing from California's Lake Tahoe (Pisani, 1974). Whatever is the historical truth, the idea was taken up again during the 1950s by the French engineer Marcel Mary (Mary, 1959). Translated into English and generalized, Mary offers the opinion that a diversion of Lake Titicaca to the Pacific Ocean by piercing at depth would provide a large head—perhaps as much as 3500 m—and would supply irrigation water to Peru's arid coast.

Trans-Andean railways, which depend on vast lengths of tunnel, are in a chaotic state at the present time. It is alleged by WIKIPEDIA that "In 2006, Ferrocarril Central Andino, work started to regauge the line from 914 mm to 1435 mm. There is also a proposal for a 21 km tunnel under the Andes" (WIKIPEDIA, accessed 29 January 2007). Such a Macro-engineering proposal is made credible by the 53 km-long English Channel Tunnel and the planning work being done by Alp-Transit Gotthard for a 57 km-long tunnel between Erstfeld and Bodio, Switzerland that, at its deepest, will be 2 km underground! The key technology that makes it possible to dig deep and long tunnels efficiently is the hard-rock Tunnel-boring Machine (TBM), a machine that excavates a tunnel by drilling out the heading to full size in one operation (Hapgood, 2004; Maidl, 2007). During early 2007, a Robbins TBM began to bore Peru's 20 km-long Olmos Transandino Project. When finished, the Olmos Transandino Project will siphon freshwater from sources higher in the Andes Mountains to a reservoir created by Limon Dam. Approximately, 2.0 billion cubic meters of freshwater will be shifted annually from the Rio Huancabamba, a tributary of the Amazon River, to the Olmos Valley in the Pacific Ocean watershed. The water will be used to irrigate 150,000 ha of farmland and will also generate ~600 MW. The Robbins TBM will have to negotiate quartz porphyry and andesite geology. If Lake Titicaca were drained at depth, at its deepest point, freshwater could be made to fall ~3500 m to the Pacific Ocean.

**3. Physics of a base-of-mountain hydropower station**

Selection of the best course of the TBM-excavated tunnel will have to be done on the basis of on-site macro-engineering and geological studies. The studies will aim to predict, in the alternative suggested tunnel courses, the influence of the rock conditions on TBM operation, the amount of lining, the site and depth of the adits to the tunnel and the extent of the definitive field studies that must be done before commencement of tunnel construction. The economic benefits from the studies can be estimated as a considerable percentage of the total construction cost of the tunnel driving. As with other modern-day macroprojects, Macro-engineering has changed from its $20^{th}$ Century incarnation; $21^{st}$ Century Macro-engineering leaders of any project must have a list of groups to be met with, environmental impact statements to be filed, national and international laws to be complied with, and public concerns to be addressed. Yet the result—a legal and financial go-ahead—if done properly, is well worth the constraints of time and direct financial hardship: hydropower technology chosen openly, democratically, and consensually, rather than being dictated. Within a range of about 250-700 m, both Francis and impulse turbine units can be used.

Ordinarily, Lake Titicaca hydropower potential would remain worthless (on a significant geographical and economical scale) in the near-term future for a number of reasons: (1) the absence of Environmental Impact Statements; (2) formidable geological and geomorphic impediments such as infamously powerful earthquakes and rugged, even jagged, incidental terrain; (3) nearly non-existent traffic infrastructure such as highways, roads and railroads; (4) high to very high initial investment costs and long-period financial pay-backs; (5) the requirement for reliable long-distance aerial weather-resistant electric transmission lines over great distances; (6) the reluctance of international money lenders to consider low-interest loans to Peru and Bolivia; (7) volatile and inconsistent national political opinions regarding priorities of national, regional and centralized or decentralized energy system development and (8) the considerable on-going development of alternative energy resources such as natural gas fields.

Furthermore, we must assume that future global climate change may instigate flexibility requirements for many existing and planned infrastructures! If, for example, the Altiplano climate becomes drier than today, then Lake Titicaca will be reduced in volume and the freshwater could be wasted. On the other hand, if the Altiplano becomes wetter, Lake Titicaca will refill faster and pose a severe damage threat to all established infrastructures surrounding that body of water that is wedged in the Andes Mountains! Our offered technical suggestion creates a situation whereby—in either instance—Peru's coast-sited population will flourish and prosper! The map of Titicaca region is shown in FIGURE 1.

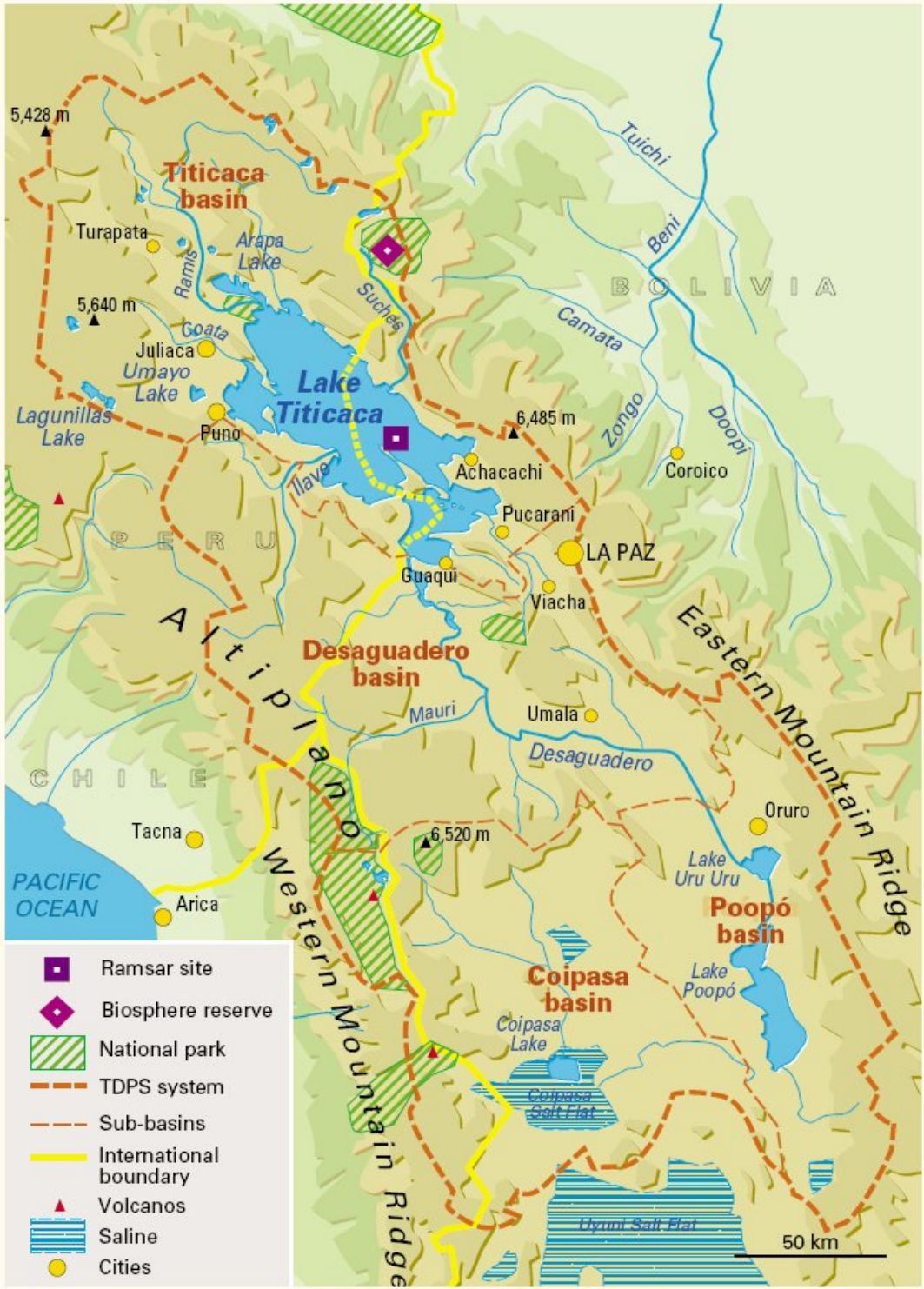

**FIGURE 1.** Lake Titicaca.

# Most relevant data about Lake Titicaca

Data on Lake Titicaca is derived from different researchers from different time periods—in other words, data must assumed to be somewhat "iffy". Here, we show the most reliable generalities about Lake Titicaca.
1. Catchment total area 58,000 km$^2$.
2. Lake Titicaca's average surface area 8,372 km$^2$.
3. Water volume 893-930 km$^3$.
4. Average depth 107 m
5. Maximum depth 283-351 m.
6. Modern Surface elevation 3812 m
7. Primary source is 27 rivers. Main influx is Rio Ramis (flow 76 m$^3$/s).
8. Effluent Rio Desaguadero (flow 20-35 m$^3$/s).
9. Annual precipitation 703 mm/year.
10. Annual evapo-transpiration 652 mm/year (173 m$^3$/s).
11. Annual discharge 281 m$^3$/s.
12. Altitude of western ridge is 4000-4200 m.
13. Lake Titicaca's level oscillation is about 1 m/year up to ~6.47 m in 29 year period.
14. Used water 9.5 m$^3$/s (include irrigation 7.4 m$^3$/s).
15. Annual fish catch is 6,327 tons.
16. Economic losses of ~$890,000/decade are due to floods.
17. Usual cost of electricity in Peru is about $3 kilowatt-hour.

Lake Titicaca extends between $14^0$ and $20^0$ South Latitude and between $66^0$ and $71^0$ West Longitude. It is 176 km long and approximately 70 km wide.

Table 21.6: Annual flow in ten control stations of Lake Titicaca and Desaguadero River

| River | Station | Average (m³/s) | Maximum (m³/s) | Minimum (m³/s) |
|---|---|---|---|---|
| Ramis | Ramis | 75.6 | 130.4 | 24.4 |
| Huancané | Huancané | 20.0 | 38.8 | 6.9 |
| Suchez | Escoma | 10.6 | 18.9 | 4.0 |
| Coata | Maravilla | 41.5 | 75.5 | 2.4 |
| Ilave | Ilave | 38.5 | 96.6 | 5.0 |
| Desaguadero | International | 35.5 | 186.5 | –3.5 |
| Desaguadero | Calacoto | 51.9 | 231.6 | 6.2 |
| Mauri | Abaroa | 4.9 | 9.8 | 2.3 |
| Caquena | Abaroa | 2.8 | 5.6 | 0.9 |
| Mauri | Calacoto | 18.6 | 31.8 | 5.7 |
| Desaguadero | Ulloma | 77.1 | 282.7 | 19.7 |
| Desaguadero | Chuquiña | 89.0 | 319.3 | 20.0 |

The main tributary of the Lake Titicaca basin is the Desaguadero River, with an average annual discharge of 89 m³/s and a maximum of 319 m³/s.

## Bolonkin-Cathcart Infrastructure Innovation

The conventional method to harness hydropower energy obtained from a high-altitude lake is by drilling a difficult-to-complete tunnel, or several such tunnels, through the enveloping hard-rock mountains. But in our case, in particular, this common method is excessively expensive, dangerous and requires long excavation period. So far, the world practice of tunneling has no experience with very long tunnels, especially those incised in hard-rock mountain geological structures.

At this time, we offer an important **technical innovation**—to put hermetic steel or prefabricated reinforced concrete tube line, both emplaced by heavy-lift helicopters, over Andes Mountains (FIGURE 2). The water pump station 2 lifts the freshwater from Lake Titicaca to the mountain ridge and into conveying hermetic tube. Then the water can safely flow to lower-elevation hydroelectric plant 4. The 90-95% energy spent for pumping will be recuperated because the lower hydroelectric plant 4 will work off of a greater water level (and higher pressure). This technical innovation decreases our unconventional facility's monetary installation cost by hundreds of times!

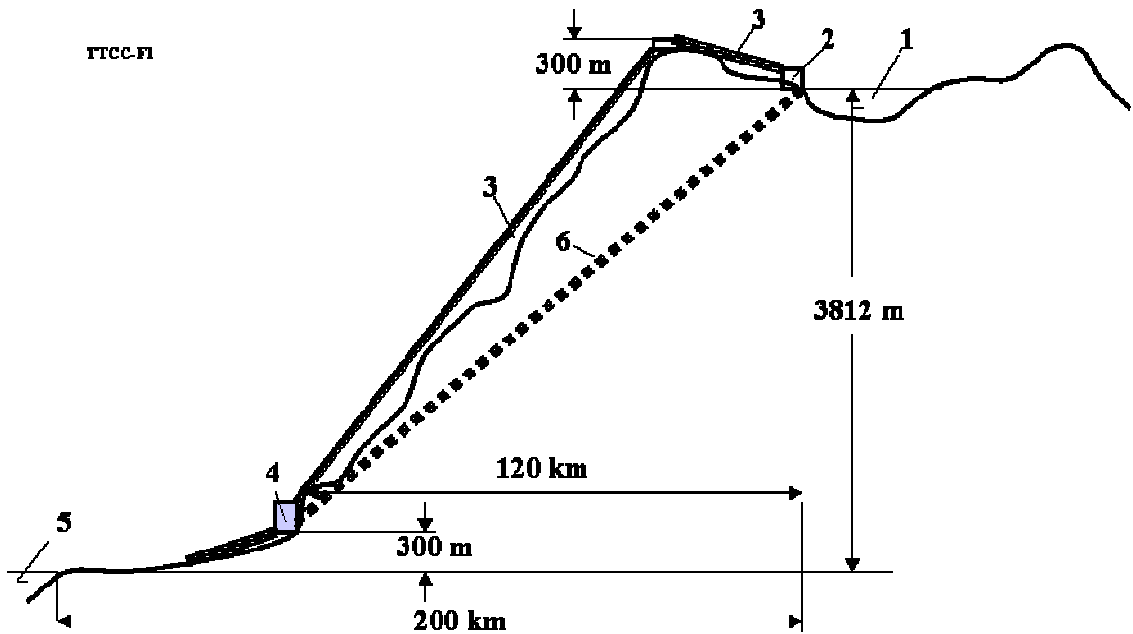

**FIGURE 2.** Sketch of proposed Lake Titicaca Electricity Generation Station. Notations: 1 – Lake Titicaca, 2 – freshwater pumping station, 3 - water tube, 4 - hydropower plant, 5 – Pacific Ocean, 6 – possible underground tunnel into a hard-rock mountain. In Peru, the shortest distance is between Puno to nearby maintain top is ~20 km, the linear distance to nearest mountain river (west of ridge) that drains to the Pacific Ocean is ~60 km and, thence, to the Pacific Ocean is ~200 km (see map in FIGURE 1). In Bolivia the shortest distance from Pucarani to Rio Zongo (after ridge) is ~50 km.

## Computations and Estimations.

1. **Estimation of the producible power** of proposed electricity station. The hydropower is computed by equation:
$$N = \eta g m H, \qquad (1)$$

where $N$ is power, W; $\eta$ is coefficient efficiency of full system, $g = 9.81$ m/s$^2$ is Earth's gravity; $m$ is water extension, m$^3$/s; $H$ is difference of water levels, m.
Computation is presented in FIGURE 3.

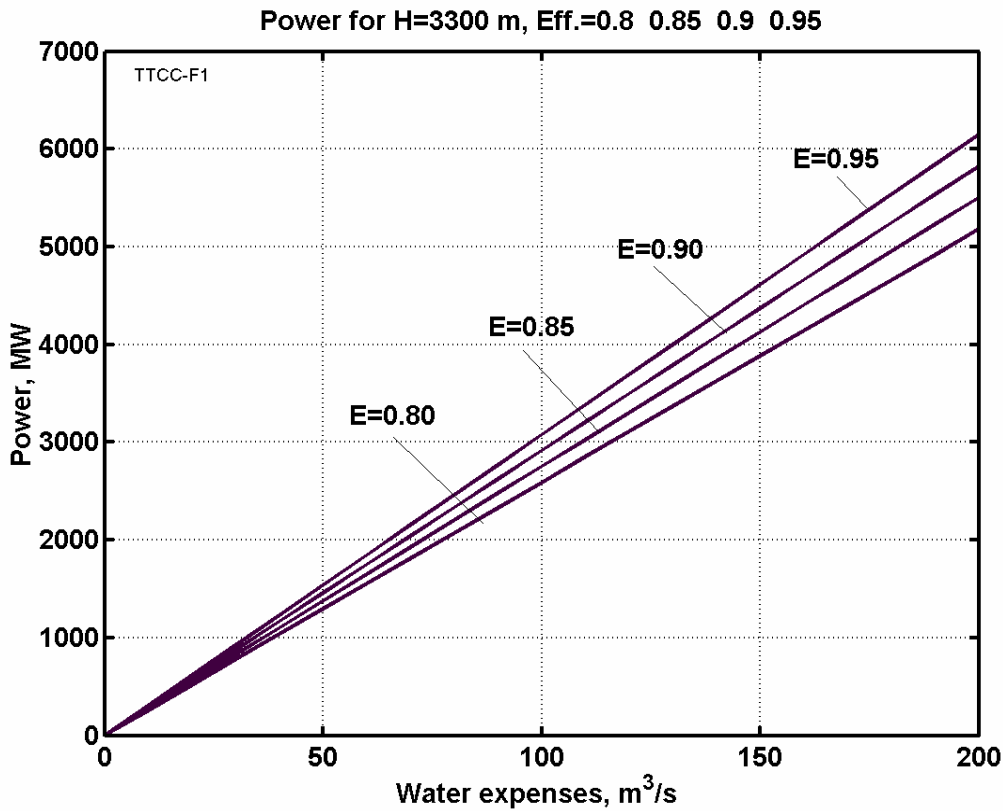

**FIGURE 3**. Full installation potential power of proposed Andean electricity plant via the water expense for different coefficients of efficiency.

As the reader can readily notice, the minimal water expense (20 m$^3$/s, without anthropogenically changing Lake Titicaca's free surface level) produces about 640 MW while the maximal expense (35 m$^3$/s, without any change of lake level) produces 1070 MW. If we agree to decrease Lake Titicaca's area by two times (100 m$^3$/s, with alteration of the lake level, FIGURE 8) then the electricity generating station can permanently produce 3060 MW. That is a very powerful electric plant, just about equaling 50% Peru's current electricity base-load!

**2. The water speed** into the hermetic tube can be estimated by equation

$$V = \frac{4m}{\pi D^2}, \qquad (2)$$

where $V$ is water speed, m/s; $m$ is water extension, m$^3$/s; $D$ is tube diameter, m.
Computation is presented in FIGURE 4. Greater tube diameter will promote less freshwater flow speed and, thus, a reduced inside water losses.

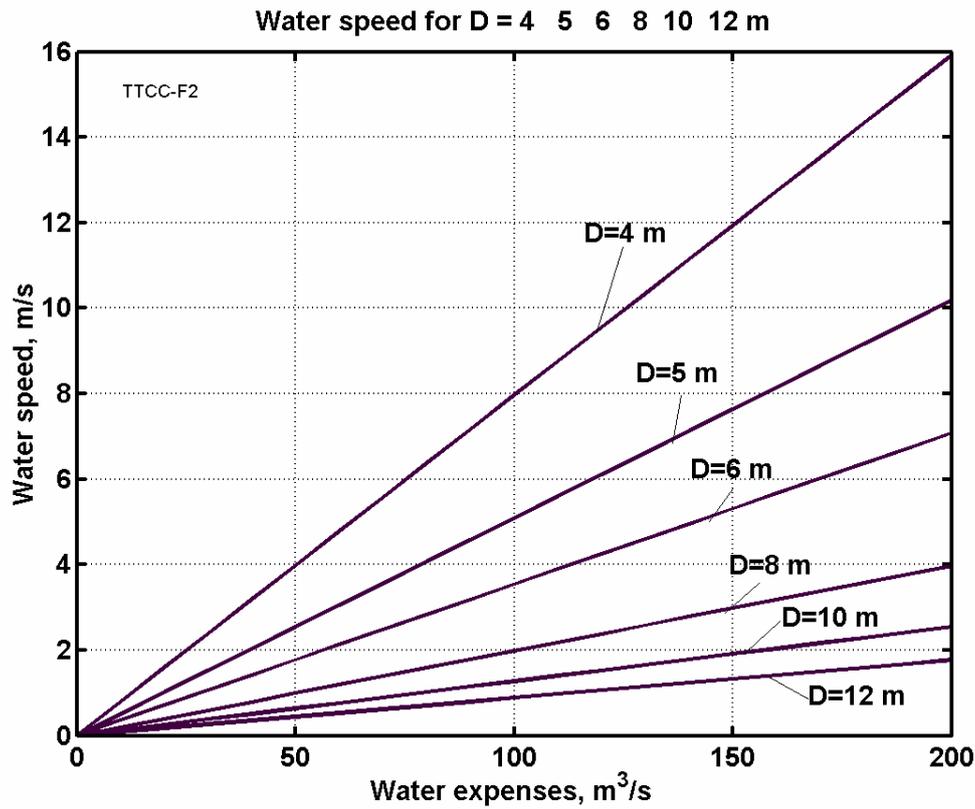

**FIGURE 4.** Freshwater speed in tube via water expenses for different tube diameters.

2. **The loss of water pressure** (in m) is computed by equation

$$h = f \frac{L}{2g} \frac{V^2}{D}, \qquad (3)$$

where $h$ is loss of water pressure, m; $f$ is friction coefficient; $L$ is length of tube, m; $V$ is water speed, m/s; $D$ is tube diameter, m.

The water friction coefficient is

$$f = \frac{0.25}{\left[\log\left(\frac{k}{3.7D} + \frac{5.14}{R_e^{0.4}}\right)\right]^2}, \qquad (4)$$

where $R_e$ is Reynolds number, $k$ is roughness.

In our case we use the reinforced concrete or steel tubes. The friction coefficient for both is approximately 0.06.

The computation of equation (3) is presented in FIGURE 5. Loss amounts to about 200 m of water pressure in a distance of 200 km or 130 m over a distance of 150 km.

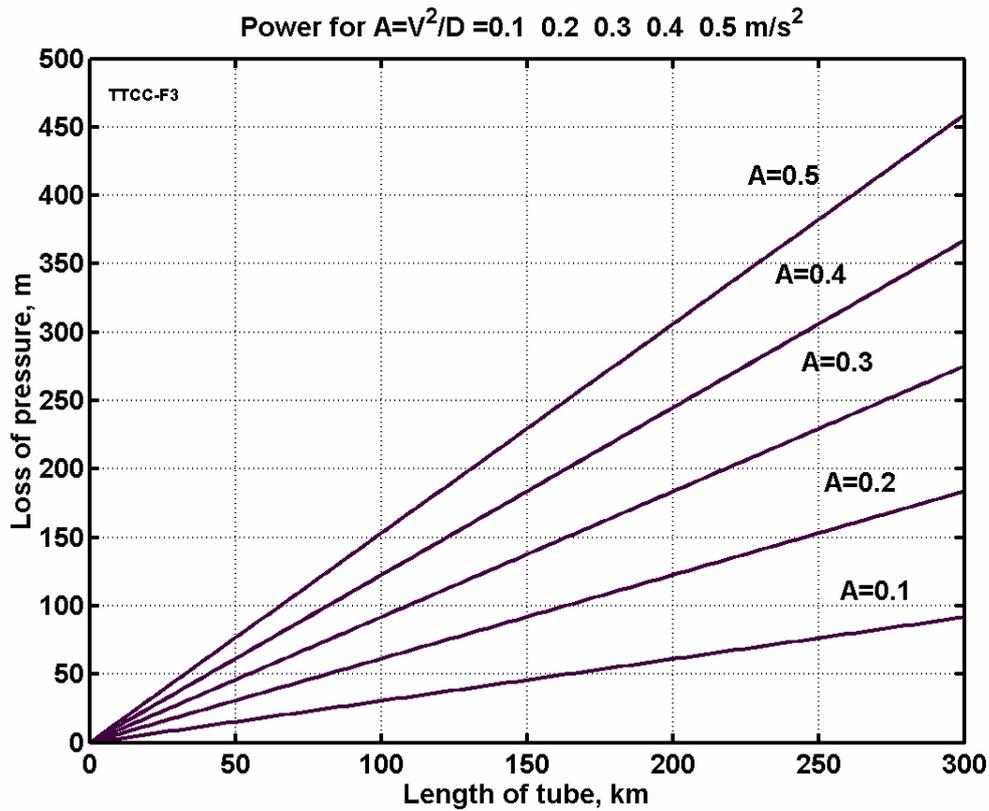

**FIGURE 5.** Loss of the water pressure into the closed tube system via the length of tube for the different ratios $A=V^2/D$, where $V$ is water speed, $D$ is hermetic tube diameter. Friction coefficient $f = 0.06$.

3. **Relative loss of water pressure**. The relative loss of the tube's water pressure may be estimated by equation

$$\bar{h} = f \frac{V^2}{2gD}, \qquad (5)$$

where $\bar{h} = h/L$ is relative loss of water pressure, m/km. The computation is presented in FIGURE 6.

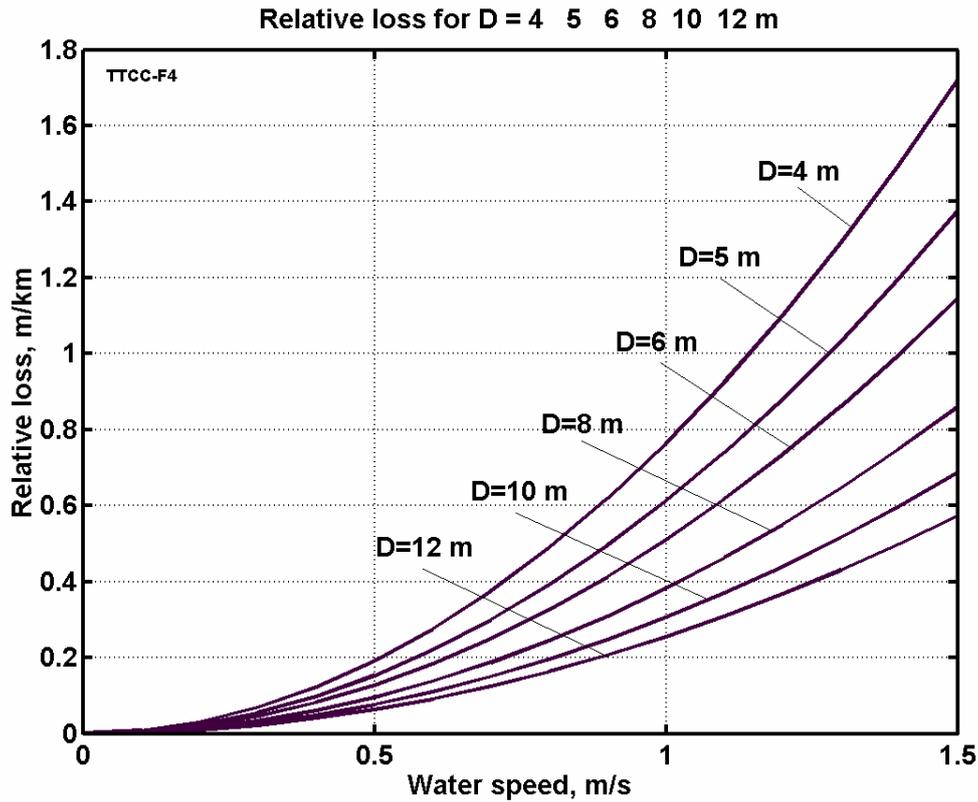

**FIGURE 6.** Relative loss (m/km) of the water pressure via the water speed for different tube diameters. Friction coefficient $f = 0.06$.

The big hydraulic tube (with a large diameter) is more expensive but such a tube has the advantage of decreasing greatly the pressure loss and increases significantly the efficiency and the power of our proposed electric-generation station for Peru located west of the Andes Mountains.

**4. Tubes**. Tubes near the lower hydropower station have high pressure (up 380 atm.). They must be made from steel or something even stronger—perhaps from some composite fiber material. Such composed material has higher maximum stress (up 600 kg/mm$^2$, steel has only 120 kg/mm$^2$) and low specific-density (1800 kg/m$^3$, steel has 7900 kg/m$^3$). It may be or become cheaper quite soon. Coefficient of safety is 3 to 5. Below, is the useful equation for computation of the tube wall thickness.

$$\delta = \frac{pD}{2\sigma}, \qquad (6)$$

where $\delta$ is tube-wall thickness, m; $p$ is water pressure, N/m$^2$; $\sigma$ is safety tensile stress, N/m$^2$. The result of computation is presented in FIGURE 7.

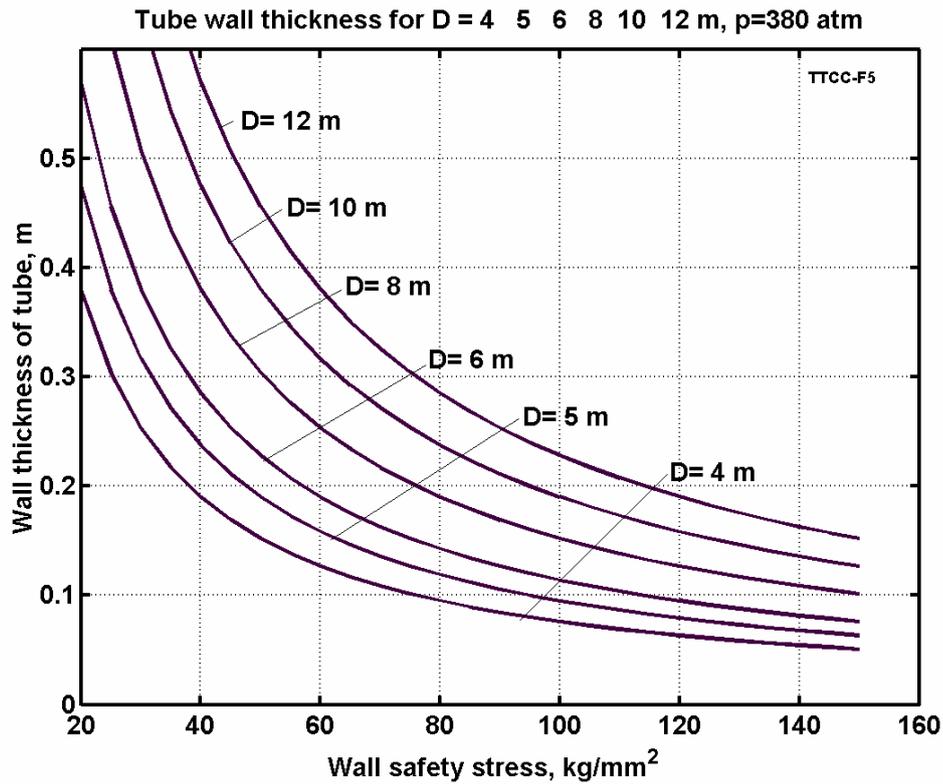

**FIGURE 7.** Tube-wall thickness via wall safety tensile stress for different tube diameters and water pressure 380 atm.

**5. Over-expenditure of water**. In case of permanent water over-expenditure, Lake Titicaca will be balanced at a new, human-selected lower free-surface water level. The lake's area decreases and evaporation also decreases. The site-specific computations for Lake Titicaca are presented in FIGURE 8. The maximum of water extension is about 200 m$^3$/s, producing a maximum electric power of about 6100 MW. That is equivalent to all electric power plants of Peru at the present time! But, eventually, Lake Titicaca will vanish as a geographic feature. (It is possible to imagine that future global climate change might eventuate in a wetter regional climate allowing the constancy of this terminal lake.) The other way is to decrease evaporation by using some special anthropogenic floating coverings that create a thin layer on the freshwater's free surface and, thereby, retard or even prohibit evaporation or, ultimately, to simply cover the water surface with an extremely pliable and thin fabric film.

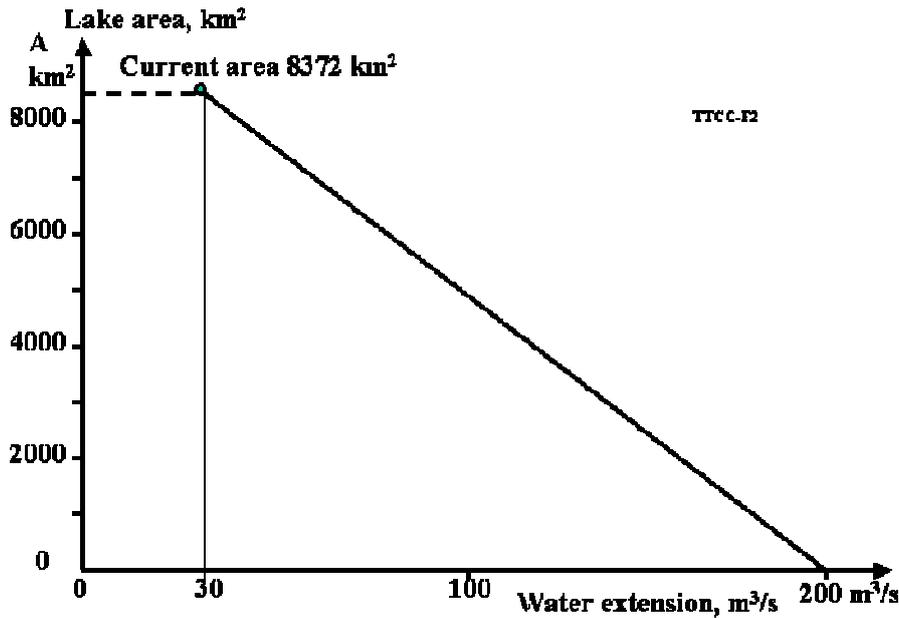

**FIGURE 8**. Decreasing of the lake's surface area via average flow through the power plant. If the average flow is less then 30 m³/s, we can decrease the loss (flow) in the Rio Desaguadero and retain Lake Titicaca's level. (Nowadays, there is a dam regulating the Rio Desaguadero.) Any permanent use of more than 30 m³/s inevitably decreases the lake's free surface area.

## Conclusion

The inexpensive hydroelectric power plant, having generation capability of 640-1000 MW may be built near Lake Titicaca without any hazardous, ugly change to Lake Titicaca's area or volume (water expense up 20-35 m³/s). For this to happen, the floodgates of the dam on the Rio Desaguadero near the International Bridge must be permanently closed. Average Rio Desaguadero flow is 70 m³/s and Rio Desaguadero will become a river of smaller total flow emptying into Lake Uru Uru, which is north of Lake Poopo.

If we choose to permanently utilize >30 m³/s then Lake Titicaca's free surface total area will decrease markedly. We can, however, receive, distribute and use more or less permanently a maximum up 6100 MW.  Lake Titicaca will vanish, but our innovative suggested hydroelectric station will still produce utilizable power. If, sometime, we find a truly economical method for halting or greatly reducing evaporation of the high-altitude lake, then it will certainly become possible to save Lake Titicaca in perpetuity and yet still harness a great quantity of electricity!  The electricity will be fed into the dispersed main load centers by the national power grid while the freshwater can be used for irrigation and urban use in Peru's desert-like coast region.

Most likely geopolitical obstacle: Bolivia may not agree to such freshwater diversion because Lake Titicaca is a shared resource managed by ALT. However, Bolivia also has a known need for low-cost electricity and, therefore, we think/believe that an amicable international legal treaty agreement is possible.

**References**

Baker, P.A. et al., "Holocene hydrologic variation at Lake Titicaca, Bolivia/Peru, and its relationship to North Atlantic climate variation", JOURNAL OF QUARTERNARY SCIENCE 207: 655-662 (2005).

Binational Autonomous Authority of Lake Titicaca, "Lake Titicaca Basin, Bolivia and Peru", Chapter 21, pages 466-480, IN THE 1$^{ST}$ UN WORLD WATER DEVELOPMENT REPORT: WATER FOR PEOPLE, WATER FOR LIFE (UNESCO, 2003).

Calaway, M.J., "Ice-cores, sediments and civilization collapse: a cautionary tale from Lake Titicaca", ANTIQUITY 79: 778-790 (2005).

Enock, C.R., "Water Power in the Andes", THE ENGINEER 106: 313 (1908).

Hapgood, Fred, "The Underground Cutting Edge: The innovators who made digging tunnels high-tech", INVENTION & TECHNOLOGY 20: 42-48 (Fall 2004).

Maidl, B. et al. HARDROCK TUNNEL BORING MACHINES (2007) 350 pages.

Markham, Clements R., "C. Reginald Enock's Journeys in Peru", THE GEOGRAPHICAL JOURNAL 25: 620-628 (June 1905).

Mary, Marcel, "Le Perou: ses resources hydroelecriques", HOUILLE BLANCHE 14: 450-456 (1959).

Morris, Arthur Stephen, RAISED FIELD TECHNOLOGY: THE RAISED FIELDS PROJECTS AROUND LAKE TITICACA (2004), page 1.

Mugnier, Clifford J., "Grids and Datums: Republic of Peru", PHOTOGRAMMETRIC ENGINEERING & REMOTE SENSING, pages 495-496 (May 2006).

Oncken, O. et al. THE ANDES: ACTIVE SUBDUCTION OROGENY (2006), 570 pages.

Pisani, Donald J., "'Why Shouldn't California Have the Grandest Aqueduct in the World?': Alexis Von Schmidt's Lake Tahoe Scheme", CALIFORNIA HISTORICAL QUARTERLY LIII: 347-360 (Winter 1974).

Revollo, Mario M., "Management issues in the Lake Titicaca and Lake Poopo system: Importance of developing a water budget", LAKES & RESERVOIRS: RESEARCH & MANAGEMENT 6: 225-229 (September 2001).

WIKIPEDIA, "Trans-Andean Railways": http://en.wikipedia.org/wiki/Trans-Andean_Railways (2007).